\RequirePackage{fix-cm}
\documentclass[smallextended]{svjour3}       % onecolumn (second format)
\smartqed  % flush right qed marks, e.g. at end of proof
\usepackage{graphicx}% Include figure files
\usepackage{dcolumn}% Align table columns on decimal point
\usepackage{bm}% bold math
\usepackage{amsmath,scalerel}
\usepackage{subfigure}% subcaptions for subfigures
\usepackage{subfigmat}% matrices of similar subfigures, aka small mulitples
\usepackage{mathtools,tikz,caption}
\usepackage{soul}
\usepackage{amsmath}
\begin{document}

\title{First-Order Approximation to the Boltzmann-Curtiss Equation for Flows 
with Local Spin%\thanks{Grants or other notes
}

\titlerunning{B-C Equations for Flows with Spin}        % if too long for running head

\author{Louis B. Wonnell         \and
        James Chen %etc.
}


\institute{L. B. Wonnell \at
              Kansas State University, Mechanical and Nuclear Engineering Department, Manhattan, KS 66502, USA \\
%              Tel.: +123-45-678910\\
%              Fax: +123-45-678910\\
%              \email{fauthor@example.com}           %  \\
           \and
           J. Chen \at
               Department of Mechanical and Aerospace Engineering,
              The State University of New York at Buffalo, Buffalo, NY 14260, USA\\
}

\date{Received: date / Accepted: date}

\maketitle

\begin{abstract}
The first-order approximation to the solution of the Boltzmann-Curtiss 
transport equation is derived. The resulting distribution function treats the 
rotation or gyration of spherical particles as an independent classical 
variable, deviating from the quantum mechanical treatment of molecular rotation 
found in the Wang Chang-Uhlenbeck equation. The Boltzmann-Curtiss equation, 
therefore, does not treat different rotational motions as separate molecular 
species. The first-order distribution function yields momentum equations for 
the translational velocity and gyration that match the form of the governing 
equations of morphing continuum theory (MCT), a theory derived from the 
approach 
of rational continuum thermomechanics. The contribution of the local rotation to 
the Cauchy stress and the viscous diffusion are found to be proportional to an 
identical expression based off the relaxation time, number density, and 
equilibrium temperature of the fluid. When gyration is 
equated to the macroscopic angular velocity, the kinetic description reduces to 
the first-order approximation for a classical monatomic gas, and the governing 
equations match the form of the Navier-Stokes equations. The relaxation time 
used for this approximation is shown to be more complex due to the additional 
variable of local rotation. The approach of De Groot and Mazur is 
invoked to give an initial approximation for the relaxation of the gyration. 
The incorporation of this relaxation time, and other physical parameters, into 
the coefficients of the governing equations provides a more in-depth physical 
treatment of the new terms in the MCT equations, allowing experimenters 
to test these expressions and get a better understanding of new coefficients in 
MCT.
\end{abstract}

\section{Introduction}
Researchers have been heavily relying on vorticity to characterize the flow physics of vortices, eddies and molecular rotations \cite{McCormack2012}. However, vorticity-based descriptions are often found inadequate or inconsistent when the vorticity field deviates from local rotations \cite{haller2005}. Those cases are considered as flows with strong local spins. It should be emphasized that spin is different from vorticity. Vorticity is defined by curl of velocity. In other words, it is a consequent motion of the translation. On the other hand, spin can be an independent motion. One could use the solar system as an example. Each planet spins on its own axis while moving on its own orbits. The translational veocities of any two planets are used to calculate their corresponding vorticity; however, such quantity does not represent the spinning of either planets.

Flows with strong local spin have been the focus of 
extensive theoretical, experimental, and numerical work for decades 
\cite{eringen1966theory,hirschfelder1964molecular,hynes1978molecular,jenkins1985kinetic,rahimi2016macroscopic,truesdell2004non}. High-speed, 
turbulent, compressible, reacting, and polyatomic gas 
flows all involve complex interactions based on strong local spin. The Wang 
Chan-Uhlenbeck equation accounts for molecular spin through the lens of quantum 
mechanics, treating each different quantum state as a separate species of 
molecule \cite{wang1964heat}. This additional rigor adds more complexity to 
the distribution function and the dynamics of the collisional integral. For 
classical physics, however, local rotation may affect the dynamics of the 
entire flow. Turbulent flows, in particular, may produce additional angular 
momentum from the smallest eddies. The rotation of these smallest eddies 
affects the energy and momentum transfer at the inertial length scales, 
requiring researchers to develop methods that capture this additional 
small-scale angular momentum. The most effective of these analytical methods 
have revealed deeper physical or mathematical characteristics to previously 
well-tested theories of fluid dynamics \cite{ahmadi1975turbulent,eringen1966theory,eringen:99spring,eringen:01spring,kirwan1967theory,stokes2012theories}.

Several different fields of research have adapted to flows with local spin, by 
either modifying classical theories or developing entirely new approaches. Meng 
et. al. constructed a thermal lattice Boltzmann model based on the ellipsoidal 
statistical Bhatnagar-Gross-Krook (ES-BGK) equation to capture dynamics 
of rarefied gas thermal flows \cite{meng2013lattice}. When these flows approach 
higher Mach numbers, the higher nonequilibrium flows were more difficult to 
capture in the transition regime without driving up computational costs. For 
hypersonic flows, Munafo et. al. proposed a Boltzmann rovibrational collisional 
coarse-grained model, which grouped internal energies associated with vibration 
and rotation into separate energy bins \cite{munafo2014boltzmann}. These 
groups of internal energies were treated as separate species. The main flow 
equation was simplified to a one-dimensional inviscid flow, also to save on 
computational resources. For polyatomic gases, theoretical approaches often 
treat local spin as an internal degree of freedom, similar to molecular 
vibration. Arima et. al. modified the approach of rational extended 
thermodynamics 
to treat the molecular and vibrational relaxation processes in polyatomic gases 
as separate processes, but included all effects of vibration and rotation in a 
separate variable denoting internal motions in the 
gas \cite{arima2017rational}. For rotation in nonequilibrium flows, Eu added 
to his generalized hydrodynamic relations \cite{eu1986kinetic,eu1998nonequilibrium} by introducing excess normal stress associated with a 
bulk 
viscosity \cite{eu2002generalized}. Myong et. al. developed computational models 
based on Eu's relations to analyze high Knudsen number, rarefied diatomic gas 
flows \cite{myong1999thermodynamically,myong2001computational,myong2004generalized}. 

The models previously discussed have typically been 
modifications to classical approaches, treating rotation in separate 
closure models. The history of theoretical 
work on flows with local spin, however, shows that much can be learned 
about these flows by challenging the assumptions behind the classical 
approach. From the perspective of statistical mechanics, Grad developed the 
generalized thermodynamic relations for nonequilibrium distributions of 
molecules \cite{grad1952statistical}. These relations were then applied to 
systems of molecules where individual molecules possessed internal rotation. 
When molecular rotation was treated as an internal variable dependent on the 
coordinates local to a molecule, the angular momentum equation's dependence on 
the linear momentum equation no longer held for nonequilibrium flows. 
Additionally, the pressure tensor become asymmetric due to the added internal 
rotation. De 
Groot considered the effects of an asymmetric pressure tensor on the production 
of entropy, linear momentum, and angular momentum  \cite{de1962north}. The 
stresses produced by the difference between internal rotation and the vorticity 
in the flow were characterized by the ``rotational 
viscosity'' \cite{de1962north}. Since this new coefficient appeared in the 
linear and angular momentum equations, the new parameter played the role of a 
coupling coefficient. Furthermore, when the material parameters were assumed to 
be homogeneous in space, the Navier-Stokes equations were recovered, with the 
rotational viscosity included \cite{de1962north}. 

Snider later generalized this work to account for more complex rotational 
motions in anisotropic fluids, or fluids where the local equilibrium properties 
depended on the presence of this local spin \cite{snider1967irreversible}. At 
the small and large scales, fundamental forces and properties of the fluid are 
recharacterized when molecules possess strong internal spin. The pressure 
tensor becomes asymmetric, and the torque caused by local spin gives rise 
to a couple stress tensor. This couple stress is not to be confused with 
Stokes' formulation of the couple stress, which emerges from the vorticity 
vector \cite{stokes1966couple}. Evans added depth to Snider and De Groot's 
work by 
calculating the transport coefficients in the linear constitutive equations 
that related stresses to deformations in the fluid \cite{evans1978transport}. 
Molecular dynamics simulations of dense polyatomic fluids between parallel 
plates produced preliminary data on the so-called ``vortex viscosity,'' which 
corresponded to De Groot's coupling coefficient, and showed how the variable 
for 
internal rotation approached the angular velocity in the classical limit 
\cite{evans1978transport}. The 
characteristic time that the local rotation approached the macroscopic angular 
velocity was referred to as the relaxation time. De Groot and Evans both 
derived an expression for the relaxation constant for constant vorticity and 
zero classical viscosity \cite{de1962north,evans1978transport}. For this 
simple case, the relaxation time was shown to be inversely proportional to the 
coupling coefficient. This analysis was the first indication that the return to 
equilibrium for these flows required a reformulation of the relaxation time. 
This concept will be treated with greater depth when the kinetic theory 
approach to polyatomic gases is discussed.

The presence of the new transport properties in multiple approaches to the 
problem of flows with local spin \cite{de1962north,evans1978transport}, and 
their relevance to the departure from classical fluids suggests the need for a 
deeper treatment of their physical meaning. As has been shown for the 
Navier-Stokes equations, the physics behind derived 
material constants can only come from an approach that achieves a 
macroscopic description of the fluid from modeling the interactions of 
individual particles. Maxwell and Boltzmann showed 
that these classical fluid descriptions could arise from the collisions of 
several particles in a monatomic gas \cite{boltzmann1878theorie,maxwell1873clerk}. 
Given enough collisions, a probability distribution function could predict how 
many particles would occupy a given point in space and possess a certain 
translational velocity. Maxwell 
and Boltzmann showed that a zeroth-order approximation of this distribution 
function demonstrates symmetries in physical and velocity space 
\cite{boltzmann1878theorie,maxwell1873clerk}. Furthermore, 
the substitution of the Maxwell-Boltzmann distribution into the
balance laws for mass, momentum, and energy yield governing equations that take 
the form of the Euler equations. The first-order approximation to 
the exact solution to the Boltzmann transport equation, in turn, yields 
equations that mirror the Navier-Stokes formulation. 

This initial work for monatomic gases expanded to fluids that 
contained particles with structure. Curtiss extended the Boltzmann transport 
equation to account for variables associated with separate internal motions 
apart from the translational velocity \cite{curtiss1981classical,curtiss1992classical,curtiss1963kinetic}. Additional rotation from 
diatomic molecules \cite{curtiss1981classical} and molecules of 
arbitrary structure \cite{curtiss1992classical} all resulted in additional 
terms to the Boltzmann transport equation, and complicated the physics behind 
collisional integrals and transport coefficients. The resulting transport 
equation became known as the Boltzmann-Curtiss tranpsort equation. From this 
bottom-up approach to flows with structure, Curtiss employed the 
Chapman-Enskog method \cite{curtiss1963kinetic} to obtain explicit expressions 
for the transport coefficients of a dilute gas mixture of rigid, non-spherical, 
symmetric-top molecules. Curtiss found that the resulting kinetic theory for 
this specific case matched the ``loaded-sphere'' formalism presented by Dahler 
and Sather \cite{sandler1965kinetic}. She et. al. generalized this approach to 
molecules with arbitrary internal degrees of freedom with intermolecular 
central potential forces \cite{she1967kinetic}, deriving a solution to 
the Boltzmann-Curtiss equation for molecules with translational and 
rotational motion. She also applied the Chapman-Enskog approach to find 
transport coefficients for this limited case \cite{she1967kinetic}. When 
well-known potentials were 
substituted in for the perturbations to the equilibrium distribution, the 
resulting coefficients compared with well-known models such as the 
``rough-sphere'' approximation. Solutions to the Boltzmann-Curtiss equation 
that accounted for nonequilibrium entropy production consistent with the second 
law of thermodynamics were introduced by Eu 
\cite{eu1986kinetic,eu1998nonequilibrium} and investigated by Myong 
\cite{myong1999thermodynamically,myong2001computational,myong2004generalized,myong2014high} for cases of high thermal nonequilibrium, where the complexity 
in 
the constitutive equations required a more in-depth theoretical and numerical 
treatment. The effects of rotational motion in Eu's solution were encapsulated 
by the rotational Hamiltonian, but the fluid was assumed to have no intrinsic 
angular momentum \cite{eu1998nonequilibrium}.

The extension of kinetic theory to polyatomic gases also required a detailed 
investigation into the process of the gas departing from and returning to an
equilibrium state. As mentioned in the statistical mechanical approach, the 
relaxation time typically gives a characteristic time for the return to 
equilibrium. In the classical approach, this relaxation time could be 
approximated as the time between two collisions. For the Boltzmann 
distribution, the only parameters needed to obtain this time were the mean free 
path and the most probable velocity value \cite{huang1987statistical}. For 
kinetic theories involving local rotation as an independent variable, the 
transfer of kinetic energy between translation and rotation needed to be 
considered. Parker et. al. focused on deriving the rotational relaxation time 
for homonuclear diatomic molecules \cite{parker1959rotational}. This 
description relied on the simplification that the translational energies 
already reached equilibrium, and that the rotational energy was initially 
unexcited. Furthermore, Parker did not consider rotation as a fundamental 
degree of freedom with its own equilibrium temperature 
\cite{parker1959rotational}. In the loaded-sphere 
approximation, Dahler et. al. established local equilibrium temperatures for both 
translation and rotation, and derived the rotational relaxation time from an 
approximation of the rate of transfer of energy between these motions 
\cite{dahler1963kinetic}. This approximation came from assuming a pair 
distribution for translation and rotation as the product of each of the two 
local Boltzmann distributions. Monchick et. al. was later able to establish 
singular relaxation times for rotation and other internal degrees of freedom 
based off of the Chapman-Enskog process to the linearized Boltzmann equation 
\cite{monchick1964small}. Determination of the relaxation time associated with 
rotation or vibration of gas molecules was determined experimentally via 
absorption of ultrasound frequencies and by measurements of heat conductivity 
\cite{carnevale1967ultrasonic,monchick1965heat}. Recently, 
molecular dynamics simulations have been performed to give better 
approximations of the rotational 
relaxation time \cite{valentini2012molecular} as well as other transport 
properties such as the shear viscosity. Still, these methods treat rotation as 
an internal degree of freedom, and evaluate how well the values match with 
classical treatments of the rotation in the flow. The effects of molecular 
rotation to the total relaxation time of a polyatomic gas still require a 
more detailed treatment. 

When rotation is treated as an internal or quantum state, the 
theoretician is challenged with the task of isolating its contribution 
from all other internal or quantum states. Distribution functions 
corresponding to each quantum state, as solutions to the Wang 
Chan-Uhlenbeck equation are formulated \cite{wang1964heat}, must be obtained 
and compared with avaliable data. This challenge is eliminated if rotation is 
treated as an explicit, independent variable, allowing for a single 
distribution function to describe the small-scale rotation in the system. 
Recently, an additional theory derived from the perspective of rational 
continnum thermomechanics (RCT) has provided governing equations for the 
mass, momenta, and energy of a fluid composed of spherical particles 
\cite{chen2017morphing,chen2011constitutive,eringen1966theory,eringen:99spring,eringen:01spring}. These equations 
start from a description of the fluid that 
deviates from any classical mechanical fluid. In the framework of MCT, the 
fluid is now posed as a morphing continuum, composed of individual spheres that 
possess intrinsic rotation as a separate, independent motion. It should not be confused between
the pioneering works by Eringen \cite{eringen1966theory} and Stokes \cite{stokes1966couple} and the current study. Both Erigen's and Stokes' formulations were deduced from 
rational continuum mechanics and thermodynamic irreversible processes. The current study was built upon the basis of kinetic theory
and statistical mechanics.

Similar to 
Snider's work \cite{snider1967irreversible}, the governing equations 
of morphing continuum theory (MCT) present a series of new coefficients 
directly related to the contribution of local spin 
to various stresses in the fluid. Chen showed that the inviscid equations of 
MCT could be derived from a zeroth-order approximation to the 
solution of the Boltzmann-Curtiss transport equation \cite{CHEN2017317}. 
The meaning of the additional coefficients in MCT, and their precise 
contribution to turbulent 
flows with local spin, has yet to be explained through kinetic theory. 
Peddieson et. al. derived dimensionless parameters \cite{peddieson:72jes} that 
produced a range of boundary layer profiles, including profiles that 
demonstrated aspects of 
turbulence. Still, the choice of the values for these parameters was arbitrary, 
with no expectation for which material constants were indispensable for the 
generation of turbulent fluctuations.
When governing equations derived from kinetic theory descriptions 
of a fluid mirror the form of similar equations derived from first principles, 
further insight into the role of new material constants arises.  
Material properties are shown to have intrinsic dependencies on other 
properties of the fluid. Furthermore, differences in the two sets of equations 
may reveal hidden 
assumptions in the mathematical approach or the need for higher orders of 
accuracy in the kinetic theory description. The focus of this paper is to apply 
the Chapman-Enskog approach from kinetic theory to extend Chen's analysis of 
fluids with spherical particles to the first-order approximation to the 
Boltzmann-Curtiss transport equation. The final form of this approximation is 
intended to give a deeper insight into the new coefficients introduced by MCT, 
and to present a kinetic description of a fluid possessing independent local 
rotation. The benefit of this approach of treating the local rotation as 
independent will be evident as the need for developing complex constitutive 
relations is avoided. These first-order equations should possess familiar 
terms from the Navier-Stokes equations, and 
introduce key terms that arise from local rotation. An 
analysis of these new equations will require a discussion of the relaxation 
time used to make this first-order approximation, due to the presence of local 
rotation. 

Several recent studies have shown that MCT governing equations can predict the 
turbulence velocity profile in a pipe flow \cite{Mehrabian2008}, plane Couette flow \cite{Alizadeh2011} and flat plate \cite{wonnell2017morphing}. These equations
have also been used for transonic  and supersonic \cite{Wonnell2018AIAA} turbulence \cite{cheikh2018morphing}. All these results are obtained through a
direct numerical simulation (DNS) with a fraction of the computational resources needed for NS-based DNS.
MCT is considered as a computationally friendly approach for direct numerical simulation (DNS). Therefore,
it is crucial to see the correlation between rational continuum thermomechanics and kinetic theory for morphing 
continua.

Section \ref{sec:background} specifies the assumptions for the fluid and 
outlines the mathematical consequences of making these assumptions. The 
distribution function, conservation equations, and balance laws will obtain a 
certain form from these assumptions. In section \ref{sec:first_order}, 
the 
first-order approximation to the distribution function is derived from the 
zeroth order 
balance laws by following the Chapman-Enskog approach. With the 
distribution function, the expressions for key stresses in the original balance 
laws are derived and discussed briefly. Then, section \ref{sec:GoverningEquations} 
derives the governing equations by substituting the stress tensors back into 
the original balance laws. A brief comparison with the Navier-Stokes and MCT 
linear momentum equations 
is done to highlight the new terms brought about by the local rotation of the 
spherical particles. In section \ref{sec:relaxationtime}, the physics 
underlying the new relaxation time is investigated. For section \ref{sec:navierstokes}, 
the equivalence between the gyration and the macroscopic
angular velocity results in a reduction of the governing equations to the 
Navier-Stokes description. Finally,
section \ref{sec:conclusion} concludes 
by remarking on the next steps for verifying and expanding the influence of 
this work to 
pressing problems for turbulent flows and other flows involving strong local 
spin.

\section{Background}\label{sec:background}
For monatomic gases composed of infinitesimal particles, any kinetic theory 
needs to track only the position and translational velocity of the particles. 
These assumptions greatly simplify the probability distribution of particles, 
as well as the transport equation used to describe the evolution of that 
distribution. When the particles are given a finite size and allowed to rotate, 
additional motions bring additional degrees of freedom to the system. If the 
angular motion of the particles is independent from the translational motion 
and 
is dependent on its orientation, then the transport equation has the form 
\cite{chen2017morphing,curtiss1992classical}:
 \begin{equation}
 \label{BoltzmannTransport}
  (\frac{\partial}{\partial t} + \frac{p_{i}}{m} \frac{\partial}{\partial 
x_{i}} + \frac{M_{i}}{I}\frac{\partial}{\partial \Phi_i})f = 
(\frac{\partial f}{\partial t})_{coll}
 \end{equation}
Here $m$ denotes the mass of a particle, $p_{i}$ represents the linear 
momentum, $M_{i}$ the angular momentum, $I$ 
the moment of inertia of a particle, and $\Phi_i$ the Euler angle with respect 
to the center of mass of the particle.

The solution $f(p_{i},\Phi_{i},x_{i},t)$ gives the probability a particular 
particle will possess the values of the given variables, and generalizes the 
motion of the system by simplifying the interactions of individual particles. 
For instance, this solution is absent of dependencies on vibrational energy or 
vibrational motion, as the dynamics of individual collisions are assumed to be 
independent of these variables. The right-hand side of equation 
\ref{BoltzmannTransport} accounts for the cumulative 
effect of collisions on the distribution. For this description, the particles 
are 
treated as spheres, so all axial orientations of the distribution are 
equivalent, i.e. independent of the Euler angle. Therefore, the 
Boltzmann-Curtiss transport equation becomes \cite{CHEN2017317,curtiss1992classical}:
 \begin{equation}
 \label{BoltzmannTransportSymm}
  (\frac{\partial}{\partial t} + \frac{p_{i}}{m} \frac{\partial}{\partial 
x_{i}})f = (\frac{\partial f}{\partial t})_{coll}
 \end{equation}
Equilibirum solutions to this equation should look similar to the 
Maxwell-Boltzmann distribution function, as the remaining terms are concerned 
with linear momentum. Still, the presence of an independent angular 
rotation, $\omega_{i}$, changes the distribution of kinetic energy of the 
particles. From Boltzmann's principle, the equilibrium solution to equation 
\ref{BoltzmannTransportSymm} can be approximated as \cite{CHEN2017317}:
\begin{align}
\label{BoltzmannCurtissFirst}
\ f^{0}(x_i, v_{i}, \omega_{i}, t) = \ n(\frac{\sqrt{mI}}{2\pi 
\theta})^3 \ exp(-\frac{m(v'_l v'_l) + I(\omega'_p \omega'_p)}{2\theta})
\end{align}
Here, the perturbed velocity, $v'_l = v_l - U_l$, for mean velocity $U_l$ and 
the perturbed gyration, $\omega'_p = \omega_p 
- W_p$ for mean gyration $W_p$, are introduced. The form of this distribution 
function differs from the classical Boltzmann 
distribution function \cite{kremer2010introduction,struchtrup2005macroscopic}, 
which assigns a $3/2$ power to the terms in front 
of the exponential. The increased exponential in equation 
\ref{BoltzmannCurtissFirst} arises due to the additional contribution to 
the momentum by the gyration, $\omega'_p$. The number density, $n$, of 
the particles is found by integrating 
the distribution function $f$ over all the perturbed variables, $\bf{v'}$ and 
$\bf{\omega'}$, which is now a six-dimensional integral:
\begin{equation}
 n = \int \int d^3v' d^3\omega' f^{0}
\end{equation}
The superscript indicates that this function only serves as a zeroth-order 
approximation to the true solution. In equation \ref{BoltzmannCurtissFirst}, 
the mean thermal energy $\theta$, mean velocity and mean gyration 
are assumed to vary slowly in time 
due to the rapid number of collisions, ensuring a rapid return to equilibrium. 
The thermal energy, $\theta = kT$, contains the Boltzmann constant $k$ and 
absolute temperature $T$. Classical kinetic approaches by Huang
\cite{huang1987statistical} and by Gupta et. al. for granular fluids 
\cite{gupta_shukla_torrilhon_2018} often group the Boltzmann constant with the 
characteristic temperature to focus on the thermal energy of the system.
The velocity and gyration perturbations represent the rapid 
fluctuations of the spheres, 
and provide the main source of any dynamics at equilibrium. Furthermore, the 
moment of inertia of a sphere can be expressed in terms of a parameter $j$ 
\cite{chen:12caf}, known as the microinertia. This parameter comes from the 
averaging 
of spatial coordinates attached to the sphere, allowing one to show that $j = 
\frac{2}{5}d^2$, where $d$ is the diameter of the sphere \cite{chen:12caf}. 
Substituting $I = mj$ into equation \ref{BoltzmannCurtissFirst} yields:
\begin{equation}
\label{BoltzmannCurtissSecond}
f^{0}(x_{i}, v_{i}, \omega_{i}, t) = n(\frac{m\sqrt{j}}{2\pi 
\theta})^3 exp(-\frac{m(v'_{l}v'_{l} + j\omega'_{p}\omega'_{p})}{2\theta})
\end{equation}
This equilibrium distribution function represents the starting point for the 
kinetic theory derivation, providing an abstract description of the system. To 
account for the evolution of the physical motion of a particle, the balance 
laws 
must be derived. The average of a quantity $A$ is here defined by the following 
expression:
\begin{equation}
\label{averagedefinition}
 \langle A \rangle = \frac{1}{n} \int \int A f(x_{i}, v_{i}, \omega_{i}, t) 
d^3v' d^3\omega'
\end{equation}
where $n$ is the number density of the particles and is found by integrating 
the distribution function $f$ over all the perturbed variables, $\bf{v'}$ 
and $\bf{\omega'}$. The mean velocity and gyration are naturally obtained 
from 
$\langle \bf{v} \rangle$ and 
$\langle \bf{\omega} \rangle$. Therefore, any balance laws governing the mean 
velocity and mean 
gyration must come by averaging the transport 
equation \ref{BoltzmannTransportSymm} for some conserved quantity $\chi 
(x_{i}, p_{i})$:
\begin{equation}
 \label{conservationrelation}
 \frac{\partial }{\partial t} \langle n\chi
\rangle + \frac{\partial}{\partial x_{i}} \langle n
\frac{p_{i}}{m}\chi
\rangle - n \langle \frac{p_{i}}{m}\frac{\partial \chi}{\partial x_{i}}
\rangle = 0
\end{equation}
Note that all potential time derivatives vanished as $\chi$ is a function of 
momentum and position alone. The collisional term emerging from the averaging 
of the right-hand side of equation \ref{BoltzmannTransportSymm} is also 
presumed to vanish, namely, $\langle \chi(x_i, p_i) (\frac{\partial f}{\partial 
t})_{coll} \rangle = 0$. Huang proved this statement for any conserved 
quantity \cite{huang1987statistical}, and his proof will be discussed in section 
\ref{sec:first_order} when the effects of collisions are discussed in 
more detail.

The balance laws come by letting $\chi$ equal the 
conserved values of mass $m$, linear momentum $m(v_{i} + 
\epsilon_{ipl}r_l\omega_p)$, angular momentum $mr_i 
r_p\omega_p$ and total energy $m(e + \frac{1}{2}v_lv_l + 
r_pr_q\omega_p\omega_q)$ where $e=\frac{1}{2}(v'_lv'_l + 
r_pr_q\omega_p'\omega_q')$ is the internal energy density \cite{CHEN2017317} and $r_i$ is the distance vector emerging from the 
center of mass of the particle. The new velocity associated with the linear 
momentum arises from the combined motion of the classical translational 
velocity, $v_i$, and the contribution of the gyration to the total velocity, 
$\epsilon_{ipl}r_l\omega_{p}$ \cite{fowlesanalytical}. The angular momentum is 
the standard expression 
involving the the cross product of the local angular velocity induced by the 
gyration, $r_p\omega_p$, and the distance vector emerging from the 
center of mass of the particle, $r_i$. The Levi-Civita 
tensor, $\epsilon_{ipq}$, is used for cross products of two vectors, 
and has the properties:
\begin{equation}
 \label{LeviCivita}
     \epsilon_{ipq}= 
\begin{cases}
    +1,& \text{if } (i,p,q) = (x,y,z), (z, x, y), or \ (y, z, x)\\
    -1,& \text{if }  (i,p,q) = (y,x,z), (z, y, x), or \ (x, z, y)\\
     0,& \text{otherwise }
\end{cases}
\end{equation}
Finally, the conserved quantity of energy contains the kinetic 
energy associated with the local angular velocity, $r_n\omega_n$, 
and adds this to the traditional translational kinetic energy. Substituting the 
conserved quantities of mass, linear momentum, angular momentum, and energy for 
$\chi$ into the conservation equation \ref{conservationrelation} yields:
\begin{align}
\label{firstordercont}
&\textbf{Continuity} \ \Big(\chi_1 = m\Big) \nonumber \\
&\frac{\partial }{\partial t}\langle mn
\rangle + \frac{\partial }{\partial x_i}\langle mnv_i
\rangle = 0 \\
\label{firstordermoment}
&\textbf{Linear Momentum} \ \Big(\chi_2 = m(v_i + 
\epsilon_{ipl}r_l\omega_p)\Big) \nonumber \\
&\frac{\partial}{\partial t}\langle mn v_i
\rangle + \frac{\partial }{\partial t}\langle mn 
\epsilon_{ipl}r_{l}\omega_{p}
\rangle + \frac{\partial}{\partial x_{l}}\langle mn v_iv_l
\rangle + 
\\ &\frac{\partial }{\partial x_s}\langle mn\epsilon_{ipl} v_s r_{l}\omega_{p}
\rangle = 0 
\nonumber \\
\label{firstorderangmoment}
&\textbf{Angular Momentum} \ \Big(\chi_3 = mr_ir_p\omega_p \Big) \nonumber \\ 
&\frac{\partial}{\partial t}\langle mn r_{i}r_{p}\omega_p
\rangle + 
\frac{\partial}{\partial x_{l}}\langle mn r_{i}r_{p}\omega_pv_l
\rangle = 0 \\
\label{firstordereenergy}
&\textbf{Energy} \ \Big(\chi_4 = m(e + \frac{1}{2}[v_lv_l + 
r_pr_q\omega_p\omega_q])\Big) \nonumber \\
&\frac{\partial}{\partial t}\langle mn e\rangle + \frac{\partial}{\partial 
x_{i}}\langle mnev_i\rangle   + \frac{\partial}{\partial x_{i}}\frac{1}{2}\langle mn 
v'_{l}v'_{l}v'_{i} + r_{p}r_{q}\omega'_{p}\omega'_{q}v'_{i}
\rangle - \\  &  
mn\langle v_{i}\frac{\partial e}{\partial x_{i}}
\rangle = 0 \nonumber
\end{align}
Here $e$ is internal energy and is already itself a mean quantity of the 
system. Since the velocity, $\bf{v}$, and gyration, $\bf{\omega}$, are separate 
coordinates, any derivative of the positional coordinate with respect to 
these variables vanishes. Letting the averages of the variables equal their 
mean values and splitting total variables into mean and fluctuating components, 
the balance laws become:

\setlength{\belowdisplayskip}{7pt} \setlength{\belowdisplayshortskip}{7pt}
\setlength{\abovedisplayskip}{7pt} \setlength{\abovedisplayshortskip}{7pt}
 \begin{align}
\label{firstordercont2}
&\textbf{Continuity} \nonumber \\
&\frac{\partial }{\partial t}\rho + \frac{\partial }{\partial 
x_l}(\rho U_l) = 0 \\
\label{firstordermoment2}
&\textbf{Linear Momentum} \nonumber \\ 
&\frac{\partial}{\partial t}(\rho U_s)  + \frac{\partial}{\partial x_l}(\rho 
U_sU_l) + \frac{\partial }{\partial x_l}(\rho \langle v'_s v'_l
\rangle + \langle \rho 
\epsilon_{spq} v'_l r_{q}\omega'_{p}
\rangle) = 0\\
\label{firstorderangmoment2}
&\textbf{Angular Momentum} \nonumber \\ 
&\frac{\partial}{\partial t}(\rho i_{sp}W_p) + 
\frac{\partial}{\partial x_{l}}(\rho i_{sp}W_pU_l) + \frac{\partial}{\partial 
x_{l}}(\rho \langle i_{sp}\omega'_p v'_l
\rangle) = 0 \\ \nonumber 
\label{firstordereenergy2}
&\textbf{Energy} \nonumber \\ 
&\frac{\partial}{\partial t}(\rho e) + \frac{\partial}{\partial 
x_{l}}(\rho eU_l)   + \frac{\partial}{\partial x_{l}}\frac{1}{2}\langle \rho 
v'_{s}v'_{s}v'_{l} + i_{pq}\omega'_{q}\omega'_{p}v'_{l}
\rangle  -  
\rho \langle v_{l}\frac{\partial e}{\partial x_{l}}
\rangle = 0 
\end{align}
Here, the properties $\langle v'\chi
\rangle = 0$ and $\langle \omega'\chi
\rangle = 0$ are employed. Additionally, the term $\langle \epsilon_{ipl} v_s 
r_{l}W_{p} \rangle = 0$ as this can be viewed as an integral of the 
fluctuating component of the total velocity 
\cite{baraff1997introduction,CHEN2017317}. Also, 
the term $i_{pq} = r_pr_q$ is used to represent the product of the coordinates, 
$r_p$, emerging from the center of mass of the particle. 
These coordinates measure the relative deformation of a particle, 
tracking how the surface varies about the center of mass.
The tensor $i_{pq}$ is related to the earlier parameter $j$, known as the 
microinertia. 
For spherical particles, $i_{pq}$ is reduced to $i_{pq}\delta_{pq} = 
i_{pp}$, which can be 
shown to equal $\frac{3j}{2}$ \cite{chen:12caf}. 
Applying this reduction to 
$i_{pq}$ the balance laws become:
 \begin{align}
\label{firstordercont3}
&\textbf{Continuity} \nonumber \\
&\frac{\partial }{\partial t}\rho + \frac{\partial }{\partial 
x_l}(\rho U_l) = 0 \\
\label{firstordermoment3}
&\textbf{Linear Momentum} \nonumber \\ 
&\frac{\partial}{\partial t}(\rho U_s)  + \frac{\partial}{\partial x_l}(\rho 
U_sU_l) + \frac{\partial }{\partial x_l}(\rho \langle v'_s v'_l
\rangle + \langle \rho 
\epsilon_{spq} v'_l r_{q}\omega'_{p}
\rangle) = 0\\
\label{firstorderangmoment3}
&\textbf{Angular Momentum} \nonumber \\ 
&\frac{\partial}{\partial t}(\frac{3\rho jW_s}{2}) + 
\frac{\partial}{\partial x_{l}}(\frac{3\rho jW_sU_l}{2}) + 
\frac{\partial}{\partial 
x_{l}}\rho\langle\frac{3 j\omega'_s v'_l}{2}
\rangle = 0 \\ \nonumber 
\label{firstordereenergy3}
&\textbf{Energy} \nonumber \\ 
&\frac{\partial}{\partial t}(\rho e) + \frac{\partial}{\partial 
x_{l}}(\rho eU_l)   + \frac{\partial}{\partial x_{l}}\frac{1}{2}\langle \rho 
v'_{s}v'_{s}v'_{l} + \frac{3j\omega'_{p}\omega'_{p}v'_{l}}{2}
\rangle  -  
\rho \langle v_{l}\frac{\partial e}{\partial x_{l}}
\rangle = 0 
\end{align}
These conservation equations feature material derivatives for the mean flow 
variables as well as gradients of products of perturbed variables. These 
perturbations are variables in the distribution function, and so can be treated 
separately. Defining these expressions in the following way:
\begin{align}
 %&\textbf{Heat} \ \textbf{Flux}\\
 \label{heatflux}
q_{\alpha} &= \frac{1}{2}\langle \rho v'_{l}v'_{l}v'_{\alpha} + 
\frac{3j\omega'_{p}\omega'_{p}v'_{\alpha}}{2}
\rangle \\
 %&\textbf{Boltzmann} \ \textbf{Stress}\\
 \label{Boltzmannstress}
t^\text{Bol}_{\alpha \beta} &= -\rho \langle v'_\alpha v'_\beta
\rangle\\
%&\textbf{Curtiss} \ \textbf{Stress}\\
\label{Curtiss}
t^\text{Cur}_{\alpha \beta} &= - \rho \langle v'_\alpha 
\epsilon_{\beta pq}r_{q}\omega'_{p}
\rangle\\
% &\textbf{Moment} \ \textbf{Stress}\\
\label{moment}
m_{\alpha \beta} &= -\rho \langle \frac{3j\omega'_\beta v'_\alpha}{2}
\rangle
\end{align}
Here, $q_{\alpha}$ denotes the heat flux, $t^\text{Bol}_{\alpha \beta}$ gives 
the Boltzmann stress, $t^\text{Cur}_{\alpha \beta}$ yields the Curtiss stress, 
and $m_{\alpha \beta}$ introduces the moment stress. Plugging these expressions 
into the balance laws gives:
\begin{align}
\label{firstordercont4}
&\textbf{Continuity} \nonumber \\ 
&\frac{\partial }{\partial t}\rho + \frac{\partial }{\partial x_l}(\rho U_l) = 
0 \\
\label{firstordermoment4}
&\textbf{Linear Momentum} \nonumber \\ 
&\frac{\partial}{\partial t}(\rho U_s)  + \frac{\partial}{\partial x_l}(\rho 
U_sU_l) - \frac{\partial }{\partial x_l}(t^{\text{Bol}}_{ls} + 
t^{\text{Cur}}_{ls}) = 
0\\
\label{firstorderangmoment4}
&\textbf{Angular Momentum} \nonumber \\ 
&\frac{\partial}{\partial t}(\rho jW_s) + 
\frac{\partial}{\partial x_{l}}(\rho jW_sU_l) - 
\frac{2}{3}\frac{\partial}{\partial 
x_{l}}(m_{ls}) = 0 \\
\label{firstordereenergy4}
&\textbf{Energy} \nonumber \\ 
&\frac{\partial}{\partial t}(\rho e) + \frac{\partial}{\partial 
x_{l}}(\rho eU_l)   + \frac{\partial q_{l}}{\partial x_{l}}  -  
\rho \langle v_{l}\frac{\partial e}{\partial x_{l}}
\rangle = 0 
\end{align}
%Indeed, the expressions \ref{heatflux}, \ref{Boltzmannstress}, 
%\ref{Curtiss}, and \ref{moment} refer to familiar stresses, i.e. equation \ref{heatflux} is the heat flux, equation \ref{Boltzmannstress} is the Boltzmann stress,
 %equation \ref{Curtiss} is the Curtiss stress and equation \ref{moment} is the moment stress. 
 At the moment, they represent only source or sink terms for 
the momentum and energy of the flow. These terms can be determined 
from the definition of the average in equation \ref{averagedefinition} using 
the
equilibrium distribution in equation \ref{BoltzmannCurtissSecond}, which 
would give a very rough approximation of how they contribute to the balance 
laws. A more thorough treatment of their contribution, however, requires the 
derivation of a distribution function that accounts for departures in the fluid 
from equilibrium. For this function, the Chapman-Enskog process is followed to 
derive a first order approximation to the solution of the Boltzmann transport 
equation \ref{BoltzmannTransportSymm}.

\section{First-Order Approximation}\label{sec:first_order}
\subsection{Distribution Function}
The right-hand side of the transport equation \ref{BoltzmannTransportSymm} 
tracks the gain or loss of particles due to collisions in some small time 
interval. For the equilibrium distribution 
function in equation \ref{BoltzmannCurtissSecond}, the assumption was made 
that 
a large number of binary collisions occurred over a short time interval, 
meaning that any deviation from equilibrium would result in a rapid return to 
equilibrium. These binary collisions affect the initial rotation and velocity 
of the particle instantaneously at the moment the particles collide. Huang 
studied these binary collisions considering molecules with only translational 
velocities \cite{huang1987statistical}. The existence of spin within molecules 
was treated through the lens of quantum mechanics, denoting different spin 
states as separate species of molecules. In order to account for 
these different 
spins, then, one would need to solve the Wang Chang-Uhlenbeck equation 
\cite{wang1964heat} for the distribution function of each of these 
molecular species, with a collisional integral that accounts for the 
cross-section calculated from the quantum states of these species. Here, the 
Boltzmann-Curtiss distribution function described in equation 
\ref{BoltzmannCurtissSecond} treats gyration as an 
additional classical variable applicable to the same molecules throughout the 
domain, thus requiring only one solution to describe the distribution of 
rotation throughout the system. Additionally, the collisional 
integral is easier to calculate since the rotational motion is 
treated as a classical motion. 

The collision rate on the right-hand side of the Boltzmann transport equation 
\ref{BoltzmannTransportSymm} is given by the following integral:
\begin{equation}
 \label{collisionintegral}
 (\frac{\partial f}{\partial t})_{coll} =  \int d^3 p_2 \ d^3p_1' \ 
d^3p_2' \ \delta^4(P_f - P_i)\left|T_{fi}\right|^2(f_2'f_1' - f_2 f_1)
\end{equation}
Here, $P_f$ and $P_i$ refer to the total final and initial momenta, 
$p_1$ and $p_2$ refer to the initial momenta of the colliding 
particles while their primed counterparts, $p_1'$ and $p_2'$ each refer to 
their respective final linear momentum. As mentioned in the previous section, 
these linear momenta contain an 
added term to the classical linear momentum, $p_i = mv_i$, found in the 
Boltzmann transport equation. Here, the Boltzmann-Curtiss linear momentum, 
$p_{i} = m(v_{i} + \epsilon_{ipl}r_l\omega_p)$, includes an additional 
contribution from the component of the local rotation moving in the direction 
of 
the translational velocity. The transition matrix $T_{fi}$ contains the 
elements 
of the operator $T(E)$ that converts the particle from its initial to final 
state in the collision. Finally 
the distribution functions $f_1$ and $f_2$ refer to the distributions of 
particles containing momenta $p_1$ and $p_2$ respectively while the primed 
distribution functions contain the final momenta values denoted by the primed 
counterparts $p_1'$ and $p_2'$. Any conserved quantity for a particle 
initiating a binary collision, $\chi$, integrated with the collision integral 
\ref{collisionintegral} vanishes. Huang proved this result by interchanging 
the momenta variables before and after the collision and integrating over 
pre-collision and post-collision linear momenta \cite{huang1987statistical}. 
When equation \ref{collisionintegral} is used 
on the right-hand side of the Boltzmann-Curtiss 
transport equation \ref{BoltzmannTransportSymm}, the Wang Chan-Uhlenbeck 
equation is obtained \cite{wang1964heat}:
\begin{equation}
 \label{WangChanUhlenbeck}
 (\frac{\partial}{\partial t} + \frac{p_{i}}{m} \frac{\partial}{\partial 
x_{i}})f = \int d^3 p_2 \ d^3p_1' \ d^3p_2' \ \delta^4(P_f - 
P_i)\left|T_{fi}\right|^2(f_2'f_1' - f_2 f_1)
\end{equation}
This treatment will look at a simplified version of this 
equation.

In observing the effect of collisions on equation 
\ref{WangChanUhlenbeck}, it is important to recognize that 
$(\frac{\partial f^{0}}{\partial t})_{coll} = 0$ for the equilibrium 
Boltzmann-Curtiss distribution function defined in equation 
\ref{BoltzmannCurtissSecond}. This result emerges from the fact that 
the coefficients in equation \ref{BoltzmannCurtissSecond} do not depend on the 
velocity $v_{i}$ \cite{huang1987statistical}. To get a good approximation of 
the collision integral \ref{collisionintegral}, higher order approximations of 
$f$ are needed. 
If $g$, the deviation from the equilibrium distribution, is defined by the expression:
\begin{equation}
 g(x_i, p_i, t) = f(x_i, p_i, t) - f^{0}(x_i, p_i, t)
\end{equation}
then the collision integral, e.g. equation \ref{collisionintegral}, can be approximated with 
the following expression:
\begin{align}
 \label{collisionintegral2}\nonumber
 \hspace{0.2in}(\frac{\partial f}{\partial t})_{coll} \ \approx \ \int & d^3 p_2 
\ d^3 p'_1 \ 
d^3 p'_2 \ \delta^4 (P_f - P_i)|T_{fi}|^2 \\
& (f^{0'}_2 g_1' - f^{0}_2 g_1 
+ g_2'f^{0'}_1 - g_2 f^{0}_1)
\end{align}
where squared terms involving $g$ have been neglected due to their presumed 
smaller magnitude in relation to $f^{0}$. Indices associated with different 
distribution functions again correspond to the initial and final distributions 
of the particles in the binary collisions. To assess the relative magnitude of 
the terms within equation \ref{collisionintegral2}, the second term on the 
right-hand side can be calculated by 
the expression:
\begin{equation}
 \label{collisionmagnitude}
 -g_{1}(\textbf{x}, \textbf{p}_1, t) \int d^3p_2 \ d^3 p'_1 \ d^3 p'_2 
[\delta^4(P_f - P_i)\left|T_{fi}\right|^2f^{0}_{2} = 
-\frac{g_1}{\tau}
\end{equation}
Here, the relaxation time constant $\tau$ incorporates all the physics associated with 
the transition from initial to final states, including the transfer of 
angular 
momentum through the new variable of gyration. A more in-depth treatment of the 
gyration and the characteristic time constants associated with its evolution 
will be given in the next section.

Given the order-of-magnitude estimate to the collision integral 
\ref{collisionintegral2}, the right-hand side of the Boltzmann transport 
equation can be given a simpler treatment with the expression:
\begin{equation}
\label{firstordercollision}
 (\frac{\partial f}{\partial t})_{coll} = -\frac{f - f^{0}}{\tau} 
= -\frac{g}{\tau}
\end{equation}
The deviation function, $g$, measures the 
probability that large numbers of particles will exit their equilibrium state 
purely through collisions. The relaxation time, $\tau$, now gives an approximation 
for the entire distribution departing from equilibrium through collisions. 
Therefore, this time constant should characterize the transition of all degrees 
of freedom to and from their equilibrium states. If the time-scale of the 
problem is reduced such that only one motion departs from equilibrium, as 
Parker considered for internal rotation \cite{parker1959rotational}, then this 
time constant can be scaled to focus on this relaxation process. If further 
approximations are needed to account for additional physics, the 
relaxation time can be expanded into a series of terms that take into account 
these additional interactions. Chen et. al. applied this approach 
\cite{chen2003extended} to generate an expression for the characteristic 
collisional time scale of turbulent eddy interactions. Such expansions have the 
benefit of incorporating multiple physical processes within one time constant, 
allowing for the interaction of rotation and translation to affect the 
relaxation of the distribution function simultaneously. 

If equation \ref{firstordercollision} is 
substituted into the transport equation \ref{BoltzmannTransportSymm}, an 
approximate form of the transport equation known as the Bhatnagar, Gross, and 
Krook (BGK) equation is obtained \cite{bhatnagar1954model}:
\begin{equation}
\label{FirstOrderTransport}
 g = -\tau (\frac{\partial}{\partial t} + v_{i} \frac{\partial}{\partial 
x_{i}})(f^{0} + g)
\end{equation}
%Since $g$ measures the probability of large numbers of particles deviating from 
%their equilibrium state, its relative magnitude to $f^{0}$ matters greatly in 
%terms of what kind of system is being described. 
For this paper, it suffices to 
show what forces and properties are influencing the mean flow when slight 
deviations to equilibrium occur. Therefore, it can be assumed that $g << 
f^{0}$, 
reducing equation \ref{FirstOrderTransport} to the form:
\begin{equation}
 \label{FirstOrderTransportSimple}
  g = -\tau (\frac{\partial}{\partial t} + v_{i} \frac{\partial}{\partial 
x_{i}})f^{0}
\end{equation}
This equation gives a formula for finding $g$ entirely in terms of derivatives 
of $f^{0}$. Still, the variables in the transport equation 
\ref{FirstOrderTransportSimple} are present in $f^{0}$ only through its 
independent variables. Therefore, to get the spatial derivates of $f^{0}$, the 
following derivatives of its independent variables are calculated:
\begin{align}	
\label{derivativerhoMCT}
\frac{\partial f^{0}}{\partial \rho}&=\frac{f^{0}}{\rho} \\
\label{derivativethetaMCT}
\frac{\partial f^{0}}{\partial 
\theta}&=-(3-\frac{m(v'^2+j\omega'^2)}{2\theta})\frac{f^{0}}{ \theta}\\
\label{derivativeUMCT}
\frac{\partial f^{0}}{\partial U_i}&=\frac{mv_i'}{\theta} 
f^{0}\\
\frac{\partial f^{0}}{\partial 
W_i}&=\frac{mj\omega_i'}{\theta} f^{0}
\end{align}
Using the chain rule, the expression for $g$ in equation 
\ref{FirstOrderTransportSimple} can be written as:
\begin{equation}
\label{gDoperatorMCT}
\begin{split}
g=-\tau 
f^{0}(\frac{1}{\rho}D(\rho)&+\frac{1}{\theta}(\frac{m(v'^2+j\omega'^2)}{
2\theta }-3)D(\theta)+ 
(\frac{mv_i'}{\theta})D(U_i)\\ &+(\frac{mj\omega_i'}{\theta})D(W_i))
\end{split}
\end{equation}
where $D(X) = (\frac{\partial}{\partial t} + v_{i}\frac{\partial}{\partial 
x_{i}})X$. The material derivatives present in equation \ref{gDoperatorMCT} 
can 
be derived from the zeroth order balance laws. To obtain the zeroth order 
approximations of the equations \ref{firstordercont4}, 
\ref{firstordermoment4}, 
\ref{firstorderangmoment4}, \ref{firstordereenergy4}, the terms related to 
the 
perturbation of the velocity and gyration are eliminated, yielding: 
\begin{equation}
\label{zeroordercontMCT}
	\frac{\partial\rho}{\partial t} + \frac{\partial \rho U_l}{\partial 
x_l} = 0
\end{equation}
\begin{equation}
\label{zeroordermomentMCT}
	\frac{\partial}{\partial t}(\rho U_s)+  \frac{\partial }{\partial 
x_{l}}(\rho U_lU_s)
=-\frac{\partial }{\partial x_s} (n\theta)
\end{equation}
\begin{equation}
\label{zeroorderangmomentMCT}
	\frac{\partial}{\partial t}(\rho jW_s) + \frac{\partial }{\partial 
x_l} 
(\rho jW_sU_l)
= 0 
\end{equation}
\begin{equation}
\label{zeroordereenergyMCT}
	\frac{\partial}{\partial t}(n\theta)+ \frac{\partial }{\partial x_l} 
(n\theta U_l)
=-\frac{n\theta}{3} \frac{\partial U_q}{\partial x_q}
\end{equation}
It should be noticed that the angular momentum equation, e.g. equation \ref{zeroorderangmomentMCT}, is decoupled with the linear momentum equation, e.g. equation \ref{zeroordermomentMCT}, at the equilibrium state. At the same time, the linear momentum equation (equation \ref{zeroordermomentMCT}) is identical to the classical Euler equation. 

From these approximations to the balance laws, the material derivatives found 
in equation \ref{gDoperatorMCT} are obtained:
\begin{align}
\label{Drho}
D(\rho)&=v_l'\frac{\partial}{\partial x_l}\rho-\rho\frac{\partial 
U_q}{\partial x_q}\\
\label{Dtheta}
D(\theta)&=v_l'\frac{\partial}{\partial x_l}\theta-\frac{1}{3}\theta 
\frac{\partial U_q}{\partial x_q}\\
\label{Dui}
D(U_i)&= v_l'\frac{\partial}{\partial x_l}U_i-\frac{1}{\rho}\frac{\partial 
}{\partial x_i} (n\theta)\\
\label{Dwn}
D(W_i) &= v_l'\frac{\partial}{\partial x_l}W_i
\end{align}
With these final expressions substituted back into equation 
\ref{gDoperatorMCT}, the 
final form of $g$ is given as:
\begin{equation}
\label{gfull}
\begin{split}
g=-\tau f^{(0)} &[\frac{1}{\rho}(v_i'\frac{\partial \rho}{\partial x_i} 
-\rho 
\frac{\partial U_i}{\partial x_i}) \\
& - (\frac{3}{\theta}- 
\frac{m(v'^2+j\omega'^2)}{2\theta^2})(v'_{i}\frac{\partial \theta}{\partial 
x_i} - 
\frac{\theta}{3} \frac{\partial U_q}{\partial x_q}) \\\
 &+  
(\frac{mv'_i}{\theta})(v_l'\frac{\partial U_i}{\partial x_l} 
-\frac{1}{\rho}\frac{\partial}{\partial x_i}(n\theta)) \\
&+ 
(\frac{mj\omega_i'}{\theta})(v_l'\frac{\partial W_i}{\partial x_l})]
\end{split}
\end{equation}
Here, the first order distribution is now expressed entirely in terms of the 
mean and perturbed flow properties. All that remains is to find the first order 
approximations to the equations \ref{heatflux}, \ref{Boltzmannstress}, 
\ref{Curtiss}, and 
\ref{moment} to obtain non-zero expressions for the missing terms in the 
first-order balance laws \ref{firstordercont3}, \ref{firstordermoment3}, 
\ref{firstorderangmoment3}, and \ref{firstordereenergy3}.
\subsection{Stresses and Heat Flux}
With the definition of the heat flux in equation \ref{heatflux}, and the 
stresses in equations \ref{Boltzmannstress}, \ref{Curtiss}, and 
\ref{moment}, 
along with the 
definition of the average in equation \ref{averagedefinition}, the zeroth and 
first-order approximations to the missing terms in the balance laws can be 
calculated. Beginning with the zeroth order approximations, the averaging is 
carried out with $f^{0}$ \cite{chen2017morphing}:
\begin{align}
\begin{split}
 \label{zeroheatflux}
q^{0}_{\alpha} &= \frac{m\rho}{2n}\int (v_l'v_l'v_\alpha' + 
j\omega'_p\omega'_pv_\alpha') (\frac{m\sqrt{j}}{2\pi\theta})^3 \\
 &exp(-\frac{m(v'_{l}v'_{l} + j\omega_{p}'\omega_{p}')}{2\theta})\ 
d^3v'd^3\omega' \\ &= 0
\end{split}
\\
\begin{split}
 \label{zeroBoltzmannstress}
t^{\text{Bol}, 0}_{\alpha \beta} &= -\rho\int 
v_\alpha'v_\beta' 
(\frac{m\sqrt{j}}{2\pi\theta})^3 \\
&exp(-\frac{m(v'_{l}v'_{l} + j\omega_{p}'\omega_{p}')}{2\theta}) d^3v' d^3 
\omega' \\ 
&= -n\theta\delta_{\alpha \beta} 
\end{split}
\\
\begin{split}
 \label{zeroCurtissstress} 
t^{\text{Cur}, 0}_{\alpha \beta} &= -\rho\epsilon_{\beta 
pq}r_{p}\int 
\omega_q'v_{\alpha}' 
(\frac{m\sqrt{j}}{2\pi\theta})^3 \\
&exp(-\frac{m(v'_{l}v'_{l} + 
j\omega_{p}'\omega_{p}')}{2\theta}) d^3v' d^3 \omega'\\ 
&= 0 
\end{split}
\\
\begin{split}
 \label{zeromomentstress}
 m^0_{\alpha \beta} &= -\frac{3\rho j}{2} \int 
\omega_{\beta}'v_{\alpha}' (\frac{m\sqrt{j}}{2\pi\theta})^3 \\
& exp(-\frac{m(v'_{l}v'_{l} + 
j\omega_{p}'\omega_{p}')}{2\theta}) d^3v' d^3 \omega' 
\\ &= 0 
\end{split}
\end{align}
Due to the functional form of $f^{0}$, those integrals possessing odd powers of 
$v'_p$ or $\omega'_l$ vanish. A key note is that the Boltzmann stress in 
equation 
\ref{zeroBoltzmannstress} yields the hydrostatic pressure $P$ from $n\theta$ 
due to the assumption that the fluid is an ideal gas. This result is expected 
as the zeroth-order Boltzmann stress should reflect the zeroth order pressure 
of an ideal gas at equilibrium.\\
For the first-order approximations to the above stresses, the definitions must 
now involve volume integrals of the first-order distribution function 
$g$:
\begin{equation}
 \label{heatvolume}
 q^{1}_{\alpha} = \frac{m\rho}{2n}\int \int (v'_l v'_l v'_\alpha + \omega_p' 
\omega_p'v_{\alpha}')g d^3v' d^3\omega' 
\end{equation}
\begin{equation}
\label{Boltzmannvolume}
t^{\text{Bol},1}_{\alpha \beta} = -\rho \int \int v'_\alpha v'_\beta g d^3 v' 
d^3 \omega'
\end{equation}
\begin{equation}
\label{Curtissvolume}
t^{\text{Cur},1}_{\alpha \beta} = -\rho \int \int \epsilon_{\beta pq}r_q 
\omega_p' v'_{\alpha} g 
d^3v' d^3\omega'
\end{equation}
\begin{equation}
\label{momentvolume}
m^{1}_{\alpha \beta} = -\frac{3\rho j}{2} \int\int \omega_\alpha'v_{\beta}' g 
d^3v' 
d^3\omega' 
\end{equation}
These volume integrals are more easily evaluated if they can be converted into 
surface integrals. Since there is no angular dependence in these integrals, the 
spherical symmetry of the integrands implies:
\begin{equation}
\label{surfacedefinition}
 \int \int G(v', \omega') d^3v' d^3\omega' = 16\pi^2 \int \int 
v'^2\omega'^2 G(v',\omega')dv' d\omega'
\end{equation}
Additionally, due to the functional form of $f^{0}$, terms involving the 
average of vector components of different indices, such as $\langle 
v'_\alpha v'_\beta \rangle$, retain 
non-zero values for the integral only when indices match. Therefore, the 
identities $\langle v'_\alpha v'_\beta
\rangle = \delta_{\alpha \beta}\frac{\langle v'^2
\rangle}{3}$ and $\langle v'_{\alpha} v'_{\beta} v'_i v_l'
\rangle = \frac{\langle v^4 \rangle}{15} (\delta_{\alpha \beta}\delta_{il} + 
\delta_{\alpha i}\delta_{\beta l} + \delta_{\alpha l}\delta_{\beta i})$ are 
employed for both the velocity and gyration variables. Applying all these 
properties to our volume integrals yields:
 \begin{align}
 \begin{split}
 \label{heatMCTsurface}
 q^1_{\alpha} = &-\Bigg[\frac{8\pi^2 m \rho \tau}{3} \int \int dv' 
d\omega' (v'^6 \omega'^2 + jv'^4\omega'^4) \\
&[-\frac{4}{\theta} + 
\frac{m(v'^2 + j\omega'^2)}{2\theta^2}] \\ &(\frac{m\sqrt{j}}{2\pi \theta})^3 
exp(-\frac{m(v'^2 + j\omega'^2)}{2\theta})\Bigg] \frac{\partial 
\theta}{\partial 
x_\alpha}  \\
&= -(4n\tau\theta)\frac{\partial \theta}{\partial x_\alpha} 
\end{split}
\\
\begin{split}
 \label{surfaceBoltzmann}
 t^{\text{Bol},1}_{\alpha \beta} &= \Bigg[\frac{16\pi^2\tau\rho}{15\theta} 
\int v'^6 
\omega'^2 (\frac{m\sqrt{j}}{2\pi\theta})^3\\
& exp(-\frac{m(v'^2 + 
j\omega'^2)}{2\theta})d\omega'dv'\Bigg]\\
&(\frac{\partial U_\alpha}{\partial 
x_\beta} + \frac{\partial U_\beta}{\partial x_\alpha} + 
\delta_{\alpha \beta}\frac{\partial U_l}{\partial x_l})  \\ &- 
\Bigg[\frac{16\pi^2 
\rho \tau}{3\theta} \int v'^4 
\omega'^2 \frac{(v'^2 + j\omega'^2)}{6} 
(\frac{m\sqrt{j}}{2\pi\theta})^3 \\
&exp(-\frac{m(v'^2 
+ j\omega^2)}{2\theta})dv' d\omega' \Bigg] \delta_{\alpha \beta} \frac{\partial 
U_l}{\partial x_l}  \\
&= n\tau\theta(\frac{\partial U_\alpha}{\partial x_\beta} 
+ \frac{\partial U_\beta}{\partial x_\alpha}) - 
\frac{n\tau\theta}{3}(\frac{\partial U_l}{\partial x_l} \delta_{\alpha \beta}) 
\end{split}
\\
\begin{split}
 \label{surfaceCurtiss}
 t^{\text{Cur},1}_{\alpha \beta} = &\Bigg[\frac{16\pi^2 
\rho\tau mj}{9\theta}(\frac{m\sqrt{j}}{2\pi\theta})^3 \int v'^4\omega'^4 \\
& exp(-\frac{m(v'^2 + j\omega'^2)}{2\theta})d\omega'dv' \Bigg] \epsilon_{\beta 
pq} r_{q} \frac{\partial W_p}{\partial x_\alpha} \\
 &= (n\tau\theta) \epsilon_{\beta pq}r_q \frac{\partial W_p}{\partial x_\alpha} 
\end{split}
\\
\begin{split}
 \label{surfacemoment}
 m^1_{\alpha \beta} = &\Bigg[\frac{48\pi^2\tau\rho j^2 m}{2\theta} \int 
\omega'^2 
v'^2 \omega_{\beta}'v_{\alpha} \omega_l' v_p'
(\frac{m\sqrt{j}}{2\pi\theta})^3 \\
&exp(-\frac{m(v'^2 + 
j\omega'^2)}{2\theta})d\omega'dv' \Bigg] \frac{\partial W_l}{\partial x_p} 
\\ &= 
(\frac{3n\tau j\theta}{2})\frac{\partial W_{\beta}}{\partial x_\alpha}
\end{split}
\end{align}
The reduced forms of these stresses appear to follow familiar patterns. The 
heat flux in equation \ref{heatMCTsurface} appears to demonstrate a direct 
proportionality relationship with the temperature gradient. The Boltzmann 
stress contains terms related to the familiar strain-rates and divergences of 
the velocity. Still, these stresses all have nonlinear dependence on the 
temperature, meaning that simplifications will have to be made before direct 
comparisons with classical fluids can occur.
\section{Governing Equations}\label{sec:GoverningEquations}
\begin{figure}[ht!]
\centering
 \includegraphics[width=.6\linewidth]{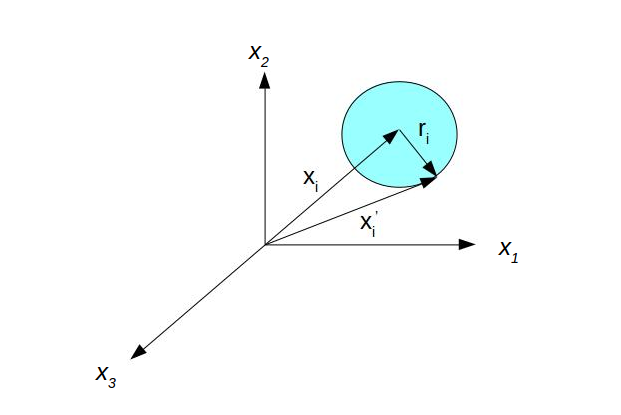}
 \caption{Illustration of the relationship between coordinates $r_i$ and $x_i$}
 \label{f:coordfigure}
\end{figure}

\begin{figure}
\centering
 \includegraphics[width=.5\linewidth]{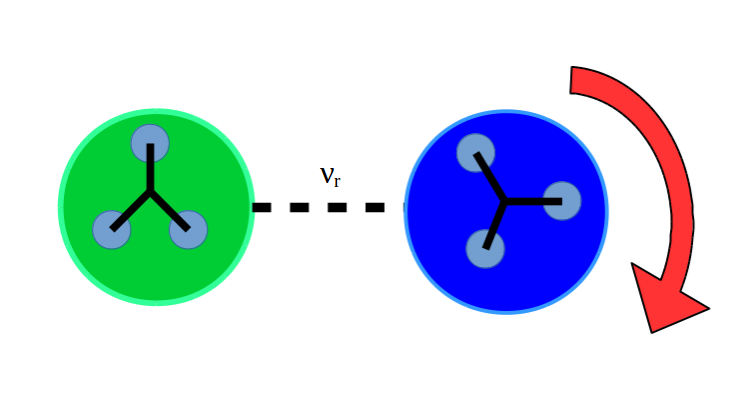}
 \caption{Diagram of the angular momentum body force $L_{s}$. Presence of 
vorticity induces gyration of left structure, with the strength of the coupling 
effect determined by $\nu_r$.}% the value of $\kappa$.}
\label{f:relrotation}
\end{figure}

Equations \ref{heatMCTsurface}, \ref{surfaceBoltzmann}, \ref{surfaceCurtiss} and \ref{surfacemoment} serve as the constitutive models for the first order approximation to the Boltzmann-Curtiss distribution and close the governing equations.  A direct substitution of the stresses found in equations \ref{heatMCTsurface}, 
\ref{surfaceBoltzmann}, \ref{surfaceCurtiss}, and \ref{surfacemoment} into 
the 
first-order balance laws \ref{firstordercont4}, \ref{firstordermoment4}, 
\ref{firstorderangmoment4}, and \ref{firstordereenergy4} yields:
\begin{align}
\label{governingCont}
\begin{split}
&\textbf{Continuity}  \\ 
&\frac{\partial }{\partial t}\rho + \frac{\partial }{\partial x_l}(\rho U_l) = 
0 
\end{split}
\\
\begin{split}
\label{governingLinMoment}
&\textbf{Linear Momentum} \\ 
&\frac{\partial}{\partial t}(\rho U_s)  + \frac{\partial}{\partial x_l}(\rho 
U_lU_s) \\
&- \frac{\partial }{\partial x_l}[-P\delta_{sl} + n\tau\theta( 
\frac{\partial U_l}{\partial x_s} + 
\frac{\partial U_s}{\partial x_l}) - 
\frac{n\tau\theta}{3}\frac{\partial U_q}{\partial x_q}\delta_{sl} ] - \\ 
&\frac{\partial}{\partial x_{l}}(n\tau\theta\epsilon_{spq}r_q 
\frac{\partial W_p}{\partial x_l}) = 0 
\end{split}
\\
\begin{split}
\label{governingAngMoment}
&\textbf{Angular Momentum} \\ 
&\frac{\partial}{\partial t}(\rho j W_s) + 
\frac{\partial}{\partial x_{l}}(\rho jW_sU_l) - \frac{\partial}{\partial 
x_{l}}[(n\tau j\theta)\frac{\partial W_s}{\partial x_l}] = 0 
\end{split}
\\
\begin{split}
\label{governingEnergy}
&\textbf{Energy} \\
&\frac{\partial}{\partial t}(\rho e) + \frac{\partial}{\partial 
x_{l}}(\rho eU_l)   - \frac{\partial}{\partial x_{l}} 
(4n\tau\theta\frac{\partial \theta}{\partial x_l})
 - \rho \langle v_{l}\frac{\partial e}{\partial x_{l}} \rangle = 0 
\end{split}
\end{align}
These equations contain derivatives of nonlinear terms and products of 
spatially varying variables. For this first-order approximation to the balance 
laws, the products of gradients of terms are presumed to vanish. Furthermore, 
equation \ref{governingAngMoment} contains a spatial derivative of the spatial
coordinate $r_{p}$ that has its origin at the center of mass of the spherical 
particle. Looking at Figure~\ref{f:coordfigure}, the expression for this 
coordinate is easily derived in terms of the Eulerian 
coordinates: $r_{i} = x'_{i} - x_{i}$. \\
Therefore:
\begin{equation}
 \frac{\partial r_{i}}{\partial x_{l}} = - \delta_{il}
\end{equation}
Clearly the derivative is zero unless the components of $x$ and $r$ are the 
same. Taking this derivative into account, removing terms associated with 
products of gradients, and allowing for the existence of body forces, the 
governing equations become:
\begin{align}
\begin{split}
\label{governingContsimple}
&\textbf{Continuity} \\ 
&\frac{\partial }{\partial t}\rho + \frac{\partial }{\partial x_l}(\rho U_l) = 
0 
\end{split}
\\
\begin{split}
\label{governingLinMomentsimple}
&\textbf{Linear Momentum} \\ 
&\frac{\partial}{\partial t}(\rho U_s)  + \frac{\partial}{\partial x_l}(\rho 
U_sU_l) + \frac{\partial P}{\partial x_s} - n\tau\theta (\frac{\partial^2 
U_s}{\partial x_l \partial x_l} + \frac{2}{3}\frac{\partial^2 U_l}{\partial x_l 
x_s}) - \\ 
& n\tau\theta\epsilon_{spq}\frac{\partial W_q}{\partial x_p} - \rho F_{s} = 0
\end{split}
\\
\begin{split}
\label{governingAngMomentsimple}
&\textbf{Angular Momentum} \\ 
&\frac{\partial}{\partial t}(\rho j W_s) + 
\frac{\partial}{\partial x_{l}}(\rho jW_sU_l) - 
n\tau j\theta \frac{\partial^2 W_s}{\partial x_l \partial x_l} - \rho L_{s} = 0 
\end{split}
\\
\begin{split}
\label{governingEnergysimple}
&\textbf{Energy} \\
&\frac{\partial}{\partial t}(\rho e) + \frac{\partial}{\partial 
x_{l}}(\rho eU_l)   - (4n\tau\theta)\frac{\partial^2 \theta}{\partial 
x_{l}\partial x_l} 
 - \rho \langle v_{l}\frac{\partial e}{\partial x_{l}}
\rangle - \rho H = 0 
\end{split}
\end{align}
In the preceding equations the body forces $\rho F_{s}$ and $\rho L_{s}$ have 
been 
introduced to account for external phenomena unrelated to the stresses 
previously introduced. Body forces for the linear momentum are easily found 
from the classical approach and require no special treatment. In the 
independent angular momentum equation, however, the factors affecting 
$\rho L_{s}$ 
are more subtle. \ref{f:relrotation} illustrates a body force created 
by 
the presence of vorticity near an individual particle. The connection between 
the two particles is symbolized by the coefficient $\nu_r$. The motion of the 
right-hand particle creates the classical rotational motion, or 
macroscopic angular velocity, which induces the local rotation of the left 
particle. The amount of 
influence the angular velocity has on the gyration is determined by the value 
of $\nu_r$. The body force disappears once the local rotation of the left 
particle equals the angular velocity, represented by half of the vorticity. De 
Groot and Mazur characterized this body force as an asymmetric pressure tensor 
\cite{de1962north}, which had a linear relationship with the difference 
between the gyration and the angular velocity:
\begin{equation}
\label{AngMomentBody}
 \rho L^{\text{interior}}_{s} = \nu_{r}(\epsilon_{spq}\frac{\partial 
U_q}{\partial 
x_p} - 
2W_{s})
\end{equation}
Here, $\nu_{r}$ is designated as the ``rotational viscosity,'' measuring the 
strength of induced gyration on a particle caused by the presence of a 
difference between its gyration and the local vorticity. This interior body 
force couples the local rotation with the translational velocity, 
ensuring that the linear momentum equation \ref{governingLinMomentsimple} and 
angular momentum equation \ref{governingAngMomentsimple} remain intertwined as 
long as the value of $\rho L^{\text{interior}}_{s}$ remains non-zero. The total 
angular momentum body 
force, $\rho L_{s}$, can be viewed as the sum of this induced interior force 
and any external body moment force, $\rho L_{s} = \rho (L^{\text{interior}}_s + 
L^{\text{exterior}}_s)$.

The continuity equation \ref{governingContsimple} is clearly the classical 
continuity equation for the mean velocity field. The deviation from classical 
kinetic theory becomes clear in the momenta equations. The 
compressible Navier-Stokes linear momentum equation, with the assumed 
satisfaction of Stokes's hypothesis, has the form:
\begin{equation}
 \label{NSlinearmoment}
 \frac{\partial}{\partial t}(\rho U_s)  + \frac{\partial}{\partial x_l}(\rho 
U_sU_l) + \frac{\partial P}{\partial x_s} - \mu \frac{\partial^2 
U_s}{\partial x_l \partial x_l} - \frac{2\mu}{3} \frac{\partial^2 U_l}{\partial 
x_l \partial x_s} - \rho F_{s} = 0
\end{equation}
Here, $\mu$ is the dynamic viscosity of the classical fluid. Comparing 
equations \ref{NSlinearmoment} and \ref{governingLinMomentsimple}, 
the formulations are very similar, with the molecular viscosity from the 
Navier-Stokes equations represented by the expression $n\tau \theta$, as is 
expected from the first-order approximation to the Boltzmann transport equation 
\cite{huang1987statistical}. The reduction of equations 
\ref{governingContsimple}, \ref{governingLinMomentsimple}, and 
\ref{governingAngMomentsimple} to the Navier-Stokes description will be 
discussed in more detail in the next section.

The new term introduced by the preceding kinetic description has the 
form, $n\tau\theta\epsilon_{spq}\frac{\partial W_q}{\partial x_p}$. Here, the 
new variable of gyration, $W_q$, produces an additional source of linear 
momentum due to its transverse gradient. A notable insight is that the 
expression in front of the gyration gradient is also $n\tau \theta$, suggesting 
that the new term may contain a coefficient similar to the 
viscosity presented in classical fluids theory. To understand the meaning and 
importance of this new term in the linear momentum equation, the linear momentum 
equation from MCT is presented \cite{wonnell2017morphing}:
 \begin{equation}
 \begin{split}
 \label{MCTlinearmoment}
\frac{\partial}{\partial t}(\rho U_s)  &+ \frac{\partial}{\partial x_l}(\rho 
U_sU_l) + \frac{\partial P}{\partial x_s} \\ &- (\lambda + 
\mu)\frac{\partial^2 U_l}{\partial x_l \partial x_s} - (\mu + 
\kappa)\frac{\partial^2 U_s}{\partial x_l\partial x_l} \\
&- \kappa 
\epsilon_{spq}\frac{\partial W_q}{\partial x_p} - \rho F_{s} = 0
\end{split}
\end{equation}
Here, $\lambda$ represents the second coefficient of viscosity and a new 
coupling coefficient, $\kappa$, is added to the total viscosity of the MCT 
fluid. Additionally, this coupling coefficient corresponds to the 
coefficient described in \ref{f:relrotation}, as it determines the 
strength of the force induced by relative rotation within the MCT fluid. This 
theory, derived from the approach of 
rational continuum thermomechanics (RCT) 
\cite{chen2011constitutive,eringen1966theory,eringen:99spring,eringen:01spring}, starts with the same picture of the fluid 
and derives 
governing 
equations from kinematic and thermodynamic principles for a fluid with 
spherical particles. Comparing equations \ref{governingLinMomentsimple} and 
\ref{MCTlinearmoment}, the term associated with the transverse gradient in the 
kinetic equation now has a counterpart term associated with the coupling 
coefficient $\kappa$. Therefore, the first-order approximation to the 
Boltzmann-Curtiss transport equation produces a linear momentum equation 
consistent with the MCT formulation. Comparisons between the expressions for 
the 
coefficients in front of identical terms in these equations will shed light 
into 
the validity of these expressions for the new coefficients in MCT.

\section{Physical Meaning of Relaxation Time}\label{sec:relaxationtime}
The simplification of the collisional term in equation 
\ref{firstordercollision} presumes that a singular relaxation time can be used 
to describe the transition from the real distribution function $f$ to the 
equilibrium distribution function $f^{0}$. Due to the extra degrees of freedom 
introduced by the local rotation of the molecules, this relaxation time cannot 
be equated directly to the case of classical fluids. Still, as Chen et. al. 
demonstrated \cite{chen2003extended}, expressions for a singular relaxation 
time can incorporate multiple processes or models involving several degrees of 
freedom. These expressions typically start from a base time constant applied
to the relaxation of the motions of the molecular motion. In the current 
treatment, this base relaxation time would apply to the gyration. 

De Groot and 
Mazur investigated the case of viscous flow in an isotropic fluid, but allowed 
for 
asymmetry in the pressure tensor. This asymmetry required for the consideration 
of an independent conservation theorem for angular momentum. Furthermore, 
pressure asymmetry generated ``internal angular momentum,'' $S_{p}$, which 
arose 
from the local angular velocity, $\omega_{p}$, of groups of particles at a 
point 
in the system. From conservation of angular momentum, De Groot and Mazur 
derived 
a balance equation for the internal angular momentum \cite{de1962north}:
\begin{equation}
\label{internalangmoment}
 \rho\frac{dS_{q}}{dt} = -2\Pi_{q}
\end{equation}
Here, $\Pi_q$ is the asymmetrical component of the pressure tensor. Internal 
angular momentum could be easily related to the angular velocity 
through $S_{q} = I\omega_{q}$, where $I$ denoted the average moment of inertia 
of the constituent particles. The asymmetric pressure tensor, however, needed a 
more nuanced treatment. By deriving relations for the conservation of internal 
energy and entropy production, De Groot and Mazur found the thermodynamic force 
associated with the asymmetric pressure tensor \cite{de1962north}. This force 
emerged from a difference between the local and classical angular velocities, 
$\omega_{s} - \frac{1}{2}\epsilon_{spq}v_{q,p}$. Invoking Curie's principle 
\cite{curie1908symetrie} regarding thermodynamic fluxes and forces, De Groot 
and Mazur derived the following relation \cite{de1962north}:
\begin{equation}
 \label{asymmPress}
 \Pi_{s} = \nu_{r}(2\omega_{s} - \epsilon_{spq}\frac{\partial v_q}{\partial 
x_p})
\end{equation}
Clearly, the asymmetric pressure 
tensor mirrors the body force found in equation \ref{AngMomentBody}, 
indicating 
that the body force of the kinetic description can be obtained from a 
consideration of thermodynamic fluxes and forces. Given this closure relation, 
the conservation of internal angular momentum in equation 
\ref{internalangmoment} became:
\begin{equation}
 \label{intangmomentumspecific}
 \frac{d\omega_s}{dt} = -\frac{2\nu_{r}}{\rho I}(2\omega_s - 
\epsilon_{spq}\frac{\partial v_q}{\partial x_p})
\end{equation}
This equation is equivalent to the kinetic angular momentum equation 
\ref{governingAngMomentsimple} with the diffusion terms eliminated. 
Therefore, the kinetic theory is shown to obtain a more general form of a 
conservation equation. For the case of initially zero local 
angular velocity and constant vorticity, the solution to equation 
\ref{intangmomentumspecific} becomes:
\begin{equation}
 \label{intspecificsolution}
 \omega_{s} = \frac{1}{2}\epsilon_{spq}\frac{\partial v_q}{\partial x_p}(1 - 
e^{-\frac{t}{\tau_o}})
\end{equation}
where the decay of the local angular velocity is characterized by a relaxation 
time constant, $\tau_o$, that has the form:
\begin{equation}
 \label{relaxtimeDeGroot}
 \tau_o = \frac{\rho I}{4\nu_{r}}
\end{equation}
Measurements of diatomic hydrogen and deuterium mixtures at $p = 1$ atm and $T 
= 77K$ by Montero et. al. give a value of $2.20 \times 10^{-8}s$ for the 
rotational relaxation time \cite{montero2014rotational}. Thus, the assumptions 
of zero initial local rotation, constant vorticity, and 
absence of external forces, leads to the derivation of a characteristic 
relaxation time that exclusively applied to internal angular momentum. These 
assumptions become relevant when the characteristic time is sufficiently 
reduced 
such that macroscale phenomena, such as the vorticity, can be approximated as 
constant compared with the evolution of local rotation. In these short time 
scales, equilibrium is achieved for the local rotation once it approaches the 
constant vorticity. Equation \ref{relaxtimeDeGroot} provides a suitable first 
approximation of the characteristic relaxation time, $\tau$, used in our 
kinetic theory 
description. De Groot's characterization of local angular velocity as the mean 
angular velocity of groups of particles matches the physical picture of our 
kinetic theory description. The addition of body forces into the governing 
kinetic theory equations can also incorporate the thermodynamic forces found in 
De Groot and Mazur's treatment. The rotational viscosity, $\nu_{r}$, has a 
counterpart through the coupling coefficient $\kappa$ %in the body force 
%equation \cref{AngMomentBody} and 
in the MCT linear and angular momentum 
equations \cite{chen2017morphing}. Therefore, numerical 
simulations of the kinetic and MCT descriptions should be able to determine the 
appropriate conditions for the use of equation \ref{relaxtimeDeGroot} in this 
first order approximation.

\section{Reduction to Navier-Stokes Equations}\label{sec:navierstokes}
%%%%%%%
\iffalse
\section{Pathway to Navier-Stokes equations}\label{sec:PathwayToNS}
The fundamental difference between MCT and NS is the description of local rotational motion. The classical NS
equations rely on Galilean-invariant vorticity ($\epsilon_{spq}\frac{\partial 
U_q}{\partial 
x_p}$) while MCT uses the new degrees of freedom, gyration ($W_{s}$). 
 When the angular equivalence is imposed on the MCT linear
momentum equation (equation \ref{governingLinMomentsimple}), it leads to
\begin{align}
\frac{\partial}{\partial t}(\rho U_j)+\frac{\partial}{\partial x_i}(\rho U_i U_j)=-\frac{\partial p}{\partial x_j}
&+\frac{7n\tau\theta}{6}\frac{\partial^2 U_i}{\partial x_i x_j}+\frac{n\tau\theta}{2}\frac{\partial^2 U_j}{\partial x_i \partial x_i} 
\end{align}
\fi
%%%%%%

The introduction of local rotation, $\omega_{s}$, as an independent variable 
has resulted in a slightly different physical picture from the classical fluids 
description shown in the Navier-Stokes equations. The angular momentum 
equation \ref{governingAngMomentsimple} is not derived from the linear 
momentum 
equation \ref{governingLinMomentsimple}, while the classical vorticity 
equation 
can only be derived from the classical linear momentum equation previously 
shown in equation \ref{NSlinearmoment}. Still, the physical picture 
from which equations \ref{governingContsimple}, 
\ref{governingLinMomentsimple}, 
\ref{governingAngMomentsimple}, and \ref{governingEnergysimple} are derived 
differs from Boltzmann's classical picture of a monatomic gas only through the 
introduction of the variable of gyration. When the gyration of a particle is 
distinct from macroscopic rotation, as defined by the angular velocity, 
$\frac{1}{2}\epsilon_{sab}\frac{\partial U_b}{\partial x_a}$, the new form of 
the linear momentum equation \ref{governingLinMomentsimple} and the 
independent 
angular momentum equation \ref{governingAngMomentsimple} can provide an 
alternative description to the classical Navier-Stokes picture.
The difference of vorticity and gyration forms an objective (frame-indifferent) description of 
rotational motion, absolute rotation, as \cite{chen2017morphing}
\begin{equation}
\Omega^\text{AR}_s = \epsilon_{sab}\frac{\partial U_b}{\partial x_a} - 2W_{s}
\end{equation}
The disappearance of absolute rotation indicates that vorticity is solely responsible
for the local rotation and the dependence of gyration vanishes. Thus, such relation is 
called as angular equivalence.
When zero absolute rotation occurs, i.e. angular equivalence, the gyration provides no new insight from the 
classical description. Therefore, the governing equations derived in previous 
sections should reduce to the Navier-Stokes equations. Setting $W_{s} = 
\frac{1}{2}\epsilon_{sab}\frac{\partial U_b}{\partial x_a}$ in the 
governing momentum equations \ref{governingLinMomentsimple} and 
\ref{governingAngMomentsimple} yields:
\begin{align}
\begin{split}
\label{reducedLinMomentsimple}
&\textbf{Reduced Linear Momentum} \\ 
&\frac{\partial}{\partial t}(\rho U_s)  + \frac{\partial}{\partial x_l}(\rho 
U_sU_l) + \frac{\partial P}{\partial x_s} - n\tau\theta (\frac{\partial^2 
U_s}{\partial x_l \partial x_l} + \frac{2}{3}\frac{\partial^2 U_l}{\partial x_l 
\partial x_s}) - 
\\ 
& n\tau\theta\epsilon_{spq}\frac{\partial }{\partial 
x_p}(\frac{1}{2}\epsilon_{sab}\frac{\partial U_b}{\partial x_a}) - \rho F_{s} = 
0
\end{split}
\\
\begin{split}
\label{reducedAngMomentsimple}
&\textbf{Reduced Angular Momentum} \\ 
&\frac{\partial}{\partial t}(\rho \epsilon_{sab}\frac{\partial 
U_b}{\partial x_a}) + 
\frac{\partial}{\partial x_{l}}(\rho \epsilon_{sab}\frac{\partial 
U_b}{\partial x_a}U_l) - \\
& n\tau \theta \frac{\partial^2}{\partial x_l \partial 
x_l}(\epsilon_{sab}\frac{\partial U_b}{\partial x_a}) - 2\rho 
L^{\text{exterior}}_s = 
0 
\end{split}
\end{align}
The common terms of the microinertia $j$ and $\frac{1}{2}$ have been eliminated 
from equation \ref{reducedAngMomentsimple}. A key observation from equation 
\ref{reducedAngMomentsimple} is the absence of the interior body force, $\rho 
L^{\text{interior}}_{s}$, described in equation \ref{AngMomentBody}. 
The difference in rotational motions necessary for the inducement of gyration 
on a particle has 
vanished, thus making $\rho L^{\text{interior}}_{s} = 0$. Meanwhile, 
equation \ref{reducedAngMomentsimple} matches the form of 
the vorticity equation, derived from the curl of the Navier-Stokes linear 
momentum equation \ref{NSlinearmoment}:
\begin{align}
 \label{NSvorticityequation}\nonumber
 \frac{\partial}{\partial t}(\rho \epsilon_{sab}\frac{\partial U_b}{\partial 
x_a} )  & + \frac{\partial}{\partial x_l}(\rho \epsilon_{sab} 
\frac{\partial U_b}{\partial x_a} U_l) \\
&- \mu \frac{\partial^2 }{\partial x_l 
\partial x_l}(\epsilon_{sab}\frac{\partial U_b}{\partial x_a}) - \rho 
\epsilon_{sab}\frac{\partial F_{b}}{\partial x_a}  = 0
\end{align}
Looking at the reduced linear momentum equation 
\ref{reducedLinMomentsimple}, 
further manipulations will show how this equation matches the classical 
picture. Using the identity for the Levi-Civita tensor 
$\epsilon_{sab}\epsilon_{spq} = \delta_{ap}\delta_{bq} - 
\delta_{aq}\delta_{bp}$ and contracting the appropriate indices, equation 
\ref{reducedLinMomentsimple} becomes:
\begin{equation}
 \label{furtherreducedLinMoment}
 \begin{split}
&\frac{\partial}{\partial t}(\rho U_s)  + \frac{\partial}{\partial x_l}(\rho 
U_sU_l) + \frac{\partial P}{\partial x_s} - n\tau\theta (\frac{\partial^2 
U_s}{\partial x_l \partial x_l} + \frac{2}{3}\frac{\partial^2 U_l}{\partial x_l 
\partial x_s}) - 
\\ 
& \frac{n\tau\theta}{2}(\frac{\partial^2 U_{p}}{\partial x_s \partial x_p} - 
\frac{\partial^2 U_s}{\partial x_q \partial x_q}) - \rho F_{s} = 
0
\end{split}
\end{equation}
Grouping together like terms yields the type II of the Navier-Stokes 
linear momentum equation:
\begin{align}
 \label{NSlinearmomentKinetic}\nonumber
\frac{\partial}{\partial t}(\rho U_s)  + \frac{\partial}{\partial x_l}(\rho 
U_sU_l) & + \frac{\partial P}{\partial x_s} - 
\frac{n\tau \theta}{2}\frac{\partial^2 U_s}{\partial x_l \partial x_l}\\
& - \frac{7n\tau \theta}{6}\frac{\partial^2 U_{p}}{\partial x_s \partial x_p} - 
\rho 
F_s = 0
\end{align}
The form of the classical momenta equations is achieved when local rotation is 
indistinguishable from macroscopic rotation. Still, the precise formulation 
found in equations \ref{NSlinearmomentKinetic} and 
\ref{reducedAngMomentsimple} 
requires a more detailed treatment. Following the classical kinetic theory formulation and Boltzmann 
distribution, it leads to the Type I of the Navier-Stokes equations \cite{huang1987statistical}:
\begin{align}
 \label{NSlinearmomentKineticTypeI}\nonumber
\frac{\partial}{\partial t}(\rho U_s)  & + \frac{\partial}{\partial x_l}(\rho 
U_sU_l) + \frac{\partial P}{\partial x_s} \\
&- 
n\tau \theta\frac{\partial^2 U_s}{\partial x_l \partial x_l} - 
\frac{n\tau \theta}{3}\frac{\partial^2 U_{p}}{\partial x_s \partial x_p} - 
\rho 
F_s = 0
\end{align}

%The coefficient in front of the diffusion 
%term for the original linear momentum equation \ref{governingLinMomentsimple} 
%is equivalent to the classical expression $\mu = n\tau \theta$ derived by Huang 
%\cite{huang1987statistical}. 
In the type II of the Navier-Stokes equations, i.e. equation 
\ref{NSlinearmomentKinetic}, the expression for the coefficient in front of 
the 
diffusion term is half that value in the Type I, i.e. equation \ref{NSlinearmomentKineticTypeI}, due to the contribution from the new term 
associated with the curl of the gyration. This term originally contained a 
coefficient that matched the form of the classical viscosity, but applied to 
the contribution of local rotation not found in the classical description. The 
temperature dependence of viscous rotational motion appears to have a slightly 
different limiting behavior as the particle rotation begins to resemble 
macroscopic motion. 

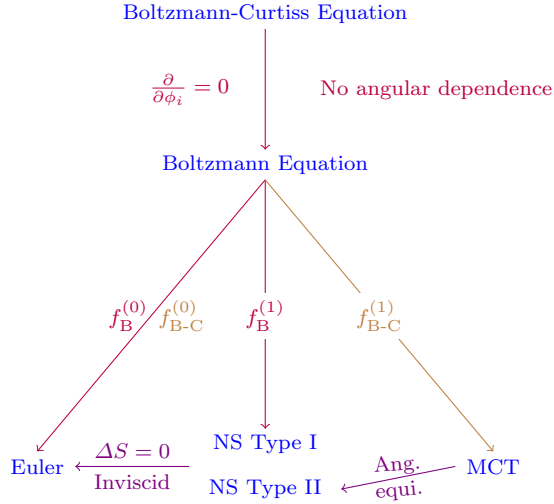
\begin{figure}
\centering
    \begin{tikzpicture}
        	\node [blue, align=center] at (0,5) {Boltzmann-Curtiss Equation};
	\draw[purple, ->] (0, 4.8) -- (0,3.2);	
	\node [purple, align=center] at (-1,4) {$\frac{\partial}{\partial \phi_i}=0$};
	\node [purple, align=center] at (2.25,4) {No angular dependence};
	\node [blue, align=center] at (0,3) {Boltzmann Equation};
	\node [blue, align=center] at (3,-1) {MCT};
	\node [blue, align=center] at (0,-0.7) {NS Type I};
	\node [blue, align=center] at (0,-1.3) {NS Type II};
	\node [blue, align=center] at (-3,-1) {Euler};
	\draw [purple, ->] (0,2.8) -- (0,-0.5);
	\node [purple, align=center, fill=white] at (0,1) {$f^{(1)}_\text{B}$};
	\draw [purple,->] (0,2.8) -- (-3,-0.8);
	\node [brown, align=center] at (-1.1,1) {$f^{(0)}_\text{B-C}$};
	\node [purple, align=center] at (-1.8,1) {$f^{(0)}_\text{B}$};
	\draw [brown, ->] (0,2.8) -- (3,-0.8);
	\node [brown, align=center, fill=white] at (1.5,1) {$f^{(1)}_\text{B-C}$}; 
	\draw [violet, ->] (2.5,-1) -- (1,-1.3);
	\node [violet, align=center, rotate=5] at (1.75,-1.15) {Ang. \\ equi.};
	\draw [violet, ->] (-1,-1) -- (-2.5,-1);
	\node [violet, align=center, below] at (-1.75,-1) {Inviscid};
	\node [violet, align=center, above] at (-1.75,-1) {$\Delta S=0$};
    \end{tikzpicture}
    \caption{Pathway from morphing continuum theory to Navier-Stokes theory}
    \label{fig:Map_NS_MCT}
\end{figure}

Figure \ref{fig:Map_NS_MCT} shows the map between morphing continuum theory and Navier-Stokes
equations from both the perspective of kinetic theory and rational continuum mechanics. 
From a kinetic point of view, Boltzmann equation can be obtained by dropping the angular
dependence in the Boltzmann-Curtiss equation. For Boltzmann equations, two different distribution
functions can be used to further deriving the conservation equations. The first one is the classical
Boltzmann distribution. When the system is at the Boltzmann distribution (zero-th order approximation), i.e. equilibrium, the Boltzmann 
equation leads to Euler's equations. Furthermore, if the system is linearly deviated from the Boltzmann distribution (first order approximation),
type I of the Navier-Stokes equation can be obtained. Similarly, if the Boltzmann -Curtiss distribution is adopted, the zero-th order approximation also leads to
the Euler's equation. It is noticed that the first order approximation of the Boltzmann-Curtiss distribution is assumed for the system, 
the transport equation leads to morphing continuum theory as presented in this study. Interestingly, one of the correlations between morphing continuum
theory and the type II of the Navier-Stokes equations is the angular motion equivalence or the vanishing absolute rotation. As discussed in the previous section, the type I and II of Navier-Stokes equation differ in the angular motion dependence of the distribution functions. This concludes the 
theoretical development and relations between classical and morphing continua.

\section{Conclusion} \label{sec:conclusion}
With the North American X-15 flying Mach 6.72 in the hypersonic regime and the 
anticipation of 
American supersonic drones by the 2030s, understanding the nonequilibrium 
phenomena in aerothermodynamic 
flows becomes a priority. Before the internal rotation and vibrations modes are 
excited, the molecular rotations first impact
the hypersonic flows. This phenomenon is considered one of the prominent 
examples for flows with local spin. Turbulence 
is another example requiring local spin. It is well-known that turbulence can 
be viewed as a tangle of vortex and eddy filaments 
\cite{jeong1995identification}. These vortex and eddy filaments are extremely 
rotational and severely impact the mean flow.
Traditionally, these rotational motions are associated with the translational 
velocities through vorticity. The vorticity, however, changes depending on the 
observer \cite{haller2005}, thus limiting its applicability to turbulence 
physics. It has been known that the classical NS theory does not yield satisfactory results 
for flow physics at high Mach number since 1920s \cite{Becker1922}. It is often believed that
the nonequilibrium phenomena in high speed flows is not captured by the first order constitutive
models from Boltzman-Maxwell distribution in NS theory. 

In the meantime, a high order continuum theory considering local spin, e.g. morphing continuum
theory \cite{chen2017morphing}, can be derived under rational continuum 
thermomechanics (RCT) or constructed from a high order distribution function, e.g. Boltzmann-Curtiss distribution \cite{CHEN2017317}. Furthermore, MCT has shown promise in analyzing 
turbulent flows \cite{cheikh2018morphing,peddieson:72jes,Silber2006,wonnell2017morphing} by considering local spin caused by eddies and vortices as 
an independent variable. The current study even show that even a first order deviation to the Boltzmann-Curtiss distribution provides satisfactory results equivalent to DSMC or Burnett equations or Super Burnett equations for hypersonic flows without any computational burdens or numerical deficiencies \cite{Mohsen2018}.
Therefore, it is 
crucial to provide the physical meaning of RCT-based continuum theories with 
the physics-based kinetic theory from the Boltzmann-Curtiss transport equation.
 
The first-order approximation to the Boltzmann-Curtiss transport equation was 
able to yield governing equations with terms corresponding to 
particular stresses in the Navier-Stokes equations and MCT. 
Furthermore, new material parameters, introduced by Chen in the 
zeroth-order approximation to fluids with spherical particles 
\cite{CHEN2017317}, received expressions based on relaxation time, number 
density, and equilibrium thermal energy. Resulting equations showed that the 
contribution of local rotation to the Cauchy stress and viscous diffusion were 
weighted equally in the kinetic description of the linear momentum equation. 
When transverse gradients in gyration disappear, the kinetic 
equation becomes the classical linear momentum equation, with the expression 
for 
the total viscosity equivalent to the result in classical fluids. It should be 
noticed that
the current formulation is at the continuum level. The rotation here refers to 
the spin of the whole molecule
and should not be confused with 
the rotation inside a molecule. The internal  rotation and vibration modes
are not within the scope and should be treated separately while these effects 
are dominant.

Further work is needed to address the proposal that the separate variables of 
gyration and 
translation each contain their own relaxation time, and that the relaxation of 
the gyration can be approximated by equation \ref{relaxtimeDeGroot}. 
Experimental efforts into flows where local rotation can be treated 
independently from translation will help answer this question, and will also 
test 
the expressions 
derived for the coefficients of the viscous diffusion and rotational Cauchy 
stress terms. The predictions of the rotational relaxation time should be 
compared with Parker \cite{parker1959rotational} and Monchick 
\cite{monchick1965heat} to determine whether the treatment of gyration as a 
non-internal, independent variable is an accurate treatment of the flow. 
Previous MCT simulations indicate \cite{cheikh2017morphing,cheikh2017energy,wonnell2016morphing,wonnell2017extension,wonnell2017morphing} that treating local rotation as an independent variable 
leads to successful modeling of turbulent fluctuations. Therefore, the 
successful derivation of the MCT equations by this kinetic description suggests 
that the independence of molecular rotation should result in realistic 
predictions of flows with strong local rotation.

\begin{acknowledgements}
This material is based upon work supported 
by the Air Force Office of Scientific Research under award number FA9550-17-1-0154. 
LBW would like to thank his coworkers Mohamad Ibrahim Cheikh and Mohamed Mohsen 
for their assistance with this work.
\end{acknowledgements}

\bibliographystyle{spmpsci}      % mathematics and physical sciences
\bibliography{references}   % name your BibTeX data base

%%%%%%%%
\iffalse

\fi
%%%%%%%%

\end{document}